\documentclass[12pt]{JHEP3}
\newcommand\dirac{\rlap/\partial}
\newcommand\Dirac{\rlap/\hskip-0.2em D}
\newcommand\indx{\hbox{index$\;$}}
\font\Nu=cmsy10 scaled 900
\def\Slash{\slash \kern-8pt}
\def\CP{{\bf CP}}
\def\longbar#1{\setbox1=\hbox{$#1$}
\setbox2=\vbox{\hrule width 0.8\wd1}
\raise0.5\ht1\hbox{${\lower\dp1\box2}\atop\box1$}}
\def\mediumbar#1{\setbox1=\hbox{$#1$}
\setbox2=\vbox{\hrule width 0.6\wd1}
\raise0.5\ht1\hbox{${\lower\dp1\box2}\atop\box1$}}  
\def\beq{\begin{equation}}
\def\enq{\end{equation}}

\title{Chiral Fermions and Spin$^c$ Structures on Matrix Approximations to Manifolds}

\author{Brian P. Dolan and C. Nash\\
Dept. of Mathematical Physics, NUI, Maynooth, Ireland\\
{\rm and}\\
School of Theoretical Physics, Dublin Institute for Advanced Studies, 
10~Burlington Rd., Dublin 8, Ireland\\
{\tt bdolan@thphys.may.ie, cnash@thphys.may.ie}\\ }

\maketitle

\begin{abstract}{
The Atiyah-Singer index theorem is investigated on various compact manifolds which admit
finite matrix approximations (``fuzzy spaces'') with a view to applications
in a modified Kaluza-Klein type approach 
in which the internal space consists of a finite number of points.  
Motivated by the chiral nature of the standard model spectrum
we investigate manifolds that do not admit spinors but do admit $Spin^c$ structures.
It is shown that, by twisting with appropriate bundles, one generation of the
electroweak sector of the standard model, including a right-handed neutrino, can
be obtained in this way from the complex projective space $\CP^2$.
The unitary Grassmannian $U(5)/\left( U(3)\times U(2) \right)$ yields a spectrum
that contains the correct
charges for the Fermions of the standard model, with varying multiplicities
for the different particle states.}

\keywords{Non-Commutative Geometry, Field Theories in Higher Dimensions, Differential and 
Algebraic Geometry}
\preprint{\tt DIAS-STP-02=08 }

\end{abstract}

\begin{document}

\section{Introduction}

Non-commutative geometry has recently come to the fore as contender for
a possible modification of physics with applications in attempts to unify
gravity and gauge theories (for reviews see \cite{Szabo}).  
Long before the current surge of 
interest via superstrings it was suggested by Connes and Lott that the standard model 
of particle physics could be derived from non-commutative geometry,
\cite{ConnesLott} \cite{Connes}.  A related concept is that of ``matrix manifolds''---these 
are a version of non-commutative geometry
in which continuous spaces with an infinite number of degrees of freedom are replaced  
with finite dimensional non-commutative matrix algebras
approximating the continuum space.  As the size of the matrices is taken to
infinity the algebra becomes commutative and the continuum space is recovered.
These algebras are often called ``fuzzy spaces'' but 
we shall refer to them as ``matrix manifolds'' in order to avoid
the negative connotations of the word ``fuzzy''.
The matrix manifold approach has much in common with generalised coherent states in quantum 
mechanics, \cite{Perelomov} \cite{Berezin} and examples of manifolds which admit a finite
matrix approximation are $S^2$ \cite{MadoreS2} \cite{PeterS2}, and more generally
$\CP^n$, \cite{FuzzyCPN} as well as unitary Grassmannians 
\begin{equation}
{U(n)\over 
U(k)\times U(n-k)}
\end{equation}
\cite{G2N}
(star products on continuous complex projective spaces and unitary Grassmannians were
constructed in \cite{Bordemann} \cite{Schirmer}).
One of the attractive features of matrix manifolds
is that they have the same symmetries as the continuum space, so a matrix
version $\left(G/H\right)_M$ of a coset space $G/H$ has all the same symmetries
of its continuous parent, despite being a finite approximation.
Matrix manifolds are also closely related to
harmonic expansions of functions on coset spaces, indeed the matrix algebras
are nothing more than cunning rearrangements of the expansion coefficients
into a matrix, and it is natural to ask if matrix manifolds might have a r\^ole to play
in Kaluza--Klein theory.  This question was investigated in \cite{Madore} and 
is one of the motivations for the present work.
Another motivation is the calculation of the spectrum
of the Dirac operator on $\CP^2_M$ in \cite{Grosse} \cite{BalCP2},
the calculation in the latter reference being built on a construction
which bears a remarkable resemblance
to the electroweak sector of the standard model.  This naturally leads one
to ask if there might be a larger matrix manifold which could incorporate the
whole standard model in its spectrum.

One of the problems with the Kaluza--Klein programme was the realisation that it
was unlikely to generate a chiral gauge theory in $4$-dimensions 
without some modification, \cite{ShelterIslandII}.  To obtain a chiral gauge
theory it seems necessary to introduce fundamental gauge fields and then
one is faced with the difficulty of anomaly cancellation, which is more
difficult in higher dimensions because there are more potentially anomalous
graphs to worry about.  The introduction of fundamental gauge fields also negates
the whole Kaluza--Klein philosophy whose aim is to derive the gauge fields purely
from a metric.  If the internal space is a matrix manifold however fundamental gauge
fields are more natural, as a matrix manifold has no simple definition of a metric,
but it does have symmetries.  To call a theory with a matrix manifold as an internal space 
a Kaluza--Klein theory is really a misnomer as all it has in common with the
usual Kaluza--Klein approach is an `internal' space with a symmetry---if a
metric is not defined there are no induced gauge fields so they must be added by hand. 
Nevertheless there is a symmetry, the symmetry of the isometries and
holonomy of the
coset space are there even at the finite level--- 
like the grin of the Cheshire cat the symmetries remain even
though the metric has gone.\footnote{For brevity we shall refer
to $G$ and $H$ as the isometry and the holonomy group, even when no metric
or connection are defined.}   
For this reason we shall continue to refer to
matrix Kaluza--Klein theory because the concept has much in common
with  continuum Kaluza--Klein theory, though there are also strong differences.

Fundamental
gauge fields are therefore natural in matrix Kaluza--Klein theories, but we
must still worry about anomalies.  To our knowledge the question of
gauge anomalies on matrix manifolds has not yet been investigated, though
chiral anomalies have been \cite{PeterS2} \cite{BalnVaidya}.  If one takes a model
consisting of $4$-dimensional Minkowski space-time with a matrix internal space
one can hope that it may be sufficient for the $4$-dimensional gauge anomalies
to cancel, without worrying about graphs with more external legs that would
be important if the internal space were continuous.  To a large extent this
is a question of dynamics on the internal space, if the dynamics
reproduces that of the continuum in the continuum limit (though it doesn't
have to if we don't want to take that limit) then these other graphs
would have to be important in the limit.
But as long as the internal space consists of a (small) number of finite
degrees of freedom it seems not unreasonable to assume that only the usual
$4$-dimensional graphs contribute to a potential anomaly.
For example the matrix manifold representing the 2-sphere 
$S^2_M$ has an approximation consisting of only
2 points.  A matrix Kaluza--Klein theory based on $S^2_M$ would look like
two copies of Minkowski space with an $SU(2)$ action on the 2 points
(much like the Higgs sector in Connes' version of the standard model).  There seems no
compelling reason to believe that such a model would exhibit a six-dimensional
gauge anomaly.

For the reasons outlined above it seems worthwhile investigating the possibility
of obtaining chiral gauge theories in $4$-dimensions from a matrix Kaluza--Klein
theory with internal space $\left(G/H\right)_M$ and fundamental gauge fields.  
The tool that we use will be standard differential geometry and
the Atiyah--Singer index theorem for the Dirac operator on continuous manifolds.
Though the aim is to apply the concepts to finite matrix geometries
it is not unreasonable to expect that the usual index theorem applies
since it makes statements about topology by counting finite data.
Indeed a Dirac operator can be defined on matrix manifolds, even though they
are finite dimensional.

The spectrum  of the Dirac operator on some specific matrix manifolds
has already been investigated, notably $S^2_M$ \cite{PeterS2} \cite{BalnVaidya}
and $\CP^2_M$, \cite{Grosse} \cite{BalCP2}.  
The construction of the spectrum on $\CP^2$ in \cite{BalCP2}
is built on a 4-dimensional reducible representation of $SU(2)\times U(1)$
which is that of electroweak sector of the standard model of particle physics, 
including a state 
with the quantum numbers of a right-handed neutrino which is a chiral zero-mode.
As is well known $\CP^2\cong SU(3)/U(2)$ is not a spin manifold, it has an obstruction to the
global definition of spinors, but coupling spinors to an appropriate background  $U(1)$ gauge field
allows spinors to be defined---a construction which is called a spin$^c$ structure
in the mathematical literature---and this gives rise to the right-handed neutrino
in \cite{BalCP2}.  
Since $\CP^2_M$ is a finite matrix
algebra approximation to continuum $\CP^2$, which captures all the topological features
of the continuum manifold, the topology is reflected
at the matrix level and the emergence of chiral spinors on $\CP^2_M$ is
a direct consequence of the Atiyah-Singer index theorem for spinors on $\CP^2$.

Since the holonomy group of $\CP^2$ is $U(2)$ the spectrum
of the Dirac operator can be decomposed into representations of $SU(2)\times U(1)$
and the representations in \cite{BalCP2} are built on that of the electroweak
sector of the standard model with the addition of a right-handed neutrino
\begin{equation} {\bf 1}_0=\hbox{\Nu V \,}_{\bf R} \qquad {\bf 1}_{-2}=e_{\bf R} \qquad 
{\bf 2}_{-1}=\left(\matrix{ \hbox{\Nu V \,}_{\bf L} \cr e_{\bf L} }\right).
\label{EW}
\end{equation}
There is a  zero-mode state in the construction of \cite{BalCP2}, 
the ${\bf 1}_0$, and its existence requires a background
Abelian `monopole' field on $\CP^2$.
In fact, as we shall see, coupling Fermions to monopole fields of higher charge
and non-Abelian background fields as well allows every state in (\ref{EW}) to be realised
as a zero-mode of the Dirac operator on $\CP^2$.
The fact that a right-handed neutrino appears naturally in the construction
is particularly appealing in view of the recent evidence for
solar neutrino oscillations \cite{SK} \cite{SNO} 
whose simplest interpretation requires a right-handed neutrino.

Another manifold which has holonomy group $U(2)$ and does not admit spinors is
\begin{equation}
{Sp(2)\over U(2)}\cong {SO(5)\over SO(3)\times SO(2)}.
\end{equation} 
This space has a finite matrix approximation and
has been proposed as a matrix version of the cotangent bundle to $S^3$ \cite{Peter}.
One might wonder
if the spectrum in (\ref{EW}) is generic for spin$^c$ structures on
manifolds with holonomy $U(2)$ and this space is a counter-example.
We shall see that a spectrum emerges which contains the correct
charges of the electroweak
sector but the Dirac operator for electron-neutrino doublet has zero index.
Nevertheless it is useful to include this as an example of a space of dimension
$2$ mod $4$ which is not spin (the significance for chiral spinors
of a distinction between spaces
of dimension $0$ mod $4$ and dimension $2$ mod $4$ was emphasised in \cite{ShelterIslandII}).

The last space which we shall examine is the matrix version
of the unitary Grassmannian
\begin{equation}
{U(5)\over U(3)\times U(2)}.
\end{equation}
This space has a finite matrix approximation and an explicit local formula for
a star-product, in terms of a finite sum of derivatives, was derived in \cite{G2N}.
It is not a spin manifold but admits a spin$^c$ structure, and so seems
a good candidate for chiral spinors.  Furthermore the holonomy group
is exactly right for the standard model, since 
\begin{equation}
{U(5)\over U(3)\times U(2)}\cong {SU(5)\over S[U(3)\times U(2)]} 
\end{equation}
and the particle spectrum of the standard model is really such that the
Fermions fall into a representation whose true group is precisely $S[U(3)\times U(2)]$,
\cite{ORaifeartaigh}.  For this space the spectrum contains one
generation of the full standard model, including a right-handed neutrino,
though the multiplicities are different for different states, some of
them having zero index.

Section 2 contains an index theorem analysis of spinors
on $\CP^2$ and reproduces the zero-mode spectrum (\ref{EW}).
Section 3 contains a discussion of the $6$-dimensional manifold $Sp(2)/U(2)$.
Section 4 analyses spin$^c$ structures, and their non-Abelian generalisations,
on the unitary Grassmannian $U(5)/(U(3)\times U(2))$ and its relation
to the standard model spectrum.
Our results are summarised in section 5.
The analysis relies on the index theorem for the Dirac
operator for various bundles over these three spaces.
The derivation of the relevant index for the cases under study is
given in four appendices, where a general discussion of spin$^c$ structures
is also given as an aid to those who may not be familiar with the construction.

\section{Chiral Fermions on ${\bf CP}^2$}

The complex projective space ${\bf CP}^2\cong SU(3)/U(2)$ 
was actively investigated in the 1980s
as an interesting candidate for an Euclidean gravitational instanton \cite{HawkingPope}.
The Euler characteristic of ${\bf CP}^2$ is $\chi=3$ and the signature is $\tau=1$.
It is not a spin manifold, there is
a global obstruction to putting spinors on this space, but one can put spinors
on it provided fundamental gauge fields are introduced and an appropriate
topologically non-trivial background gauge field is introduced. 
This fact was used in \cite{HawkingPope} to construct a ``generalised
spin structure'', where spinors with an Abelian charge move in the
field of the K\"ahler 2-form on ${\bf CP}^2$, which is somewhat analogous
to a monopole field on ${\bf CP}^1\cong S^2$.

The holonomy group of ${\bf CP}^2$ is 
\begin{equation}
U(2)\subset SO(4)\cong {SU(2)\times SU(2)\over {\bf Z}_2}.
\end{equation}
If spinors could be defined this would be lifted to $SU(2)\times U(1)\subset Spin(4)\cong
SU(2)\times SU(2)$, and the two different chiralities of Weyl spinors would transform
under the different factors of $SU(2)\times U(1)$ as, for example,
\begin{equation}
\psi_+ = {\bf 1}_1 + {\bf 1}_{-1}\qquad \hbox{and} \qquad \psi_-= {\bf 2}_0,
\end{equation}
where the subscript denotes the $U(1)$ charge.  But since spinors 
cannot be defined globally
(cf. appendix \ref{ap:CP2}), the spinor bundle does not exist.  This can be cured by introducing a $U(1)$
gauge field with non-trivial topology and 
correlating the charge 
with that of the $U(1)$ subgroup of $Spin(4)$.
Mathematically, on a complex manifold $X$,  we take the square root of the canonical line bundle $K$,
as described in appendix \ref{ap:spinc}, and tensor 
it with the the spin bundle 
$S(X)$.  Neither of these bundles exists separately but $S(X)\otimes K^{-1/2}$ 
does.  In fact, if $L$ is a {\it generating line bundle} 
(cf. the appendix) with $\int_{S^2}c_1(L)=-1$, where $S^2$ is a
non-trivial two sphere embedded in $X$,\footnote{This is ambiguous if 
$H^2(X; {\bf Z})$ has dimension greater than one, but in all the 
examples we shall consider in this paper $H^2(X; {\bf Z})$
is one dimensional and this integral is uniquely defined.}  then 
$S(X)\otimes L^p$ is a well defined bundle for any half-integral $p$.
For $\CP^2$ it is shown in appendix \ref{ap:CP2} that
\begin{equation}
S(X)\otimes L^p=\wedge^{0,*}TX\otimes K^{1/2}\otimes L^p=
\wedge^{0,*}TX\otimes L^{-q}, \quad X=\CP^2
\end{equation}
where $q=-p-{3\over 2}$, since the canonical line bundle for
${\bf CP}^2$ is given by $K=L^3$.

The net  number of zero modes depends on $q$ and for $\CP^2$ 
is given in (\ref{eq:CPq}) 
of appendix \ref{ap:CP2} as 
\begin{equation}
\nu={1\over 2}(q+1)(q+2).
\end{equation}

In fact $q$ can be interpreted as the effective $U(1)$ charge.
The charge is not $p$ because there
is a contribution from the angular momentum associated with the spinor bundle
$S(X)$.
To evaluate the charge
we use a general argument concerning spinor bundles over complex manifolds.
We define the $U(1)$ charge, which will be identified
later with the hypercharge $Y$, using
the Chern character of the generating line bundle
raised to the appropriate power, in this case
$L^{-q}$, by taking a non-trivial $S^2$ embedded in the
manifold $X$ and defining
\begin{equation}
q=\int_{S^2}ch(L^{-q})=-q\int_{S^2}ch(L)\qquad\hbox{since}\quad \int_{S^2}c_1(L)=-1.
\end{equation}

For comparison with the usual charge assignments of the
standard model below, we rescale this by $2/3$ to $Y=2q/3$.
For $q=0$ for example $\nu=1$ and, identifying positive 
chirality with right-handed spinors, this
would appear as a neutral right-handed particle:
a right-handed neutrino $\hbox{\Nu V\,}_{\bf R}$.
A spinor with $q=-3$ also has $\nu=1$, so would be right-handed, with $Y=-2$: 
the right-handed electron, $e_{\bf R}$.

If a fundamental $SU(2)$ gauge field is added with the spinors taken to
be $SU(2)$ doublets then spinors can be obtained 
from the bundle $\wedge^{0,*}T\CP^2\otimes F\otimes L^{-q}$, 
where $F$ is the rank $2$ vector
bundle defined by $F\oplus L=I^3$ ($I^3$ denoting a trivial rank 3 bundle).  
The structure group of $F$ is $U(2)$
corresponding to a $SU(2)\times 
U(1)$-gauge field.  
In fact $F$ is associated to the principal $U(2)$ bundle induced by the coset construction
\begin{equation}
\matrix{U(2)&\longrightarrow & SU(3) \cr && \downarrow \cr &&\  {\bf CP}^2\ .\cr}
\end{equation}
The Dirac index for $\wedge^{0,*}T\CP^2\otimes F\otimes L^{-q}$ 
is derived in appendix \ref{ap:CP2}
and is given by (\ref{eq:CPFq})
\begin{equation}
\nu=(q+1)(q+3).
\label{Findex}
\end{equation} 
Zero modes would give rise to chiral $SU(2)$ doublets.

The $U(1)$ charge is now calculated as the Chern character $ch(F\otimes L^{-q})$
evaluated on a topologically non-trivial $S^2$ embedded in ${\bf CP}^2$,
the result is $2q+1$.  As the Chern character involves tracing over
a $2\times 2$ matrix the $U(1)$ generator is ${(2q+1)\over 2}{\bf 1}$, where ${\bf 1}$ is the
$2\times 2$ identity matrix, so the individual charges are $q+{1\over 2}$.
Re-scaling by $2/3$, as above,
gives $Y={2q+1\over 3}$.
In particular $q=-2$ yields $Y=-1$ with $\nu=-1$ and, identifying positive chirality 
with right-handed
particles, we get a single generation of 
a left-handed doublet  with charge $-1$, the
electron-neutrino doublet.

So we can obtain a single generation of the electroweak sector
of the standard model from ${\bf CP}^2$ by taking $SU(2)$ singlets with $q=0$
and $q=-3$
($\nu=+1$) and a single $SU(2)$ doublet with $q=-2$ ($\nu=-1$),
that is
\begin{equation} {\bf 1}_0=\hbox{\Nu V \,}_{\bf R} \qquad {\bf 1}_{-2}=e_{\bf R} \qquad 
{\bf 2}_{-1}=\left(\matrix{ \hbox{\Nu V \,}_{\bf L} \cr e_{\bf L} }\right),
\qquad\nu>0\quad\hbox{right-handed}.
\label{SM}\end{equation}

\section{Chiral Fermions on $Sp(2)/U(2)$}

\label{sec:Sp2}
As an example of a six-dimensional space which does not admit a spin
structure, but does admit a $Spin^c$ structure, consider
$Sp(2)/U(2)$.
This space has Euler characteristic $\chi=4$.
In fact 
\begin{equation}
{Sp(2)\over U(2)}\cong {SO(5)\over SO(3)\times SO(2)}
\end{equation}
and this space admits a matrix approximation.
The spinor bundle does not exist but a $Spin^c$ structure
can be defined using $S(X)\otimes L^p$, with $L$ the generating line bundle and
$p$ half-integral.  
The canonical line bundle is related to the generating line bundle by $K=L^3$
(see appendix \ref{ap:Sp2}) so that 
\begin{equation}
S(X)\otimes L^p=
\wedge^{0,*}TX\otimes L^{-q},\quad X={Sp(2)\over U(2)}
\end{equation}
where $q=-p-{3\over 2}$.

The Dirac index of this bundle is derived in appendix \ref{ap:Sp2}
and is given in (\ref{eq:Spq}):
\begin{equation}
\nu={1\over 6}(2q+3)(q+1)(q+2).
\end{equation}

The zero-modes will give rise to particles in 4-dimensions whose $U(1)$
charge is $q$, which we re-scale by $2/3$ to bring it line with
the usual standard model conventions below.
so, for example, $q=-3$ gives a single generation of negative chirality
particles with charge $-2$ while $q=0$ would give a single
generation of positive chirality neutral particles.

As before we can also couple the Fermions to 
a fundamental $SU(2)$ gauge field by introducing a rank 2 vector  bundle $F$
associated to the principal bundle
\begin{equation}\matrix{U(2)&\longrightarrow& Sp(2)\cr & &\downarrow \cr &&Sp(2)/U(2) }\end{equation}
with structure group $U(2)$.
It is shown in appendix \ref{ap:Sp2} that the index of 
$\wedge^{0,*}TX\otimes F\otimes L^{-q}$
is now
\begin{equation}
\nu={2\over 3}q(q+1)(q+2).
\end{equation}
The Chern character $ch(F\otimes L^{-q})$ evaluates to 
$2q+1$ 
on a non-trivial $S^2$.  Again this is the trace of a $2\times 2$ matrix and the individual
states have charge $q+{1\over 2}$
which is rescaled by $2/3$ to give the $U(1)$
charge as $Y={2q+1\over 3}$.  
For example $q=1$ gives $Y=1$ and $\nu=4$ and thus four copies of positive
chirality doublets while $q=-2$ gives $Y=-1$ and $\nu=0$.

We can try to get the electroweak charges from this construction.
For example interpreting
positive chirality as left-handed the singlets would be the right-handed
electron $e_{\bf R}$ and a left-handed anti-neutrino 
$(\overline{\hbox{\Nu V}}\,)_{\bf L}$.  But the doublets 
with $Y=1$ would have to have negative chirality to fit with the standard model
(the right-handed positron and anti-neutrino) and $\nu$ is positive.
If we interpret positive chirality as right-handed, the doublet could
the positron--anti-neutrino doublet 
$\left(\matrix{(\overline{\hbox{\Nu V}}\,)_{\bf R}\cr
(\overline e)_{\bf R}\cr}\right)$, but then the singlet with $Y=-2$
has the wrong chirality to be the right-handed electron.

On the other hand
choosing a doublet with $q=-2$ giving $Y=-1$, in addition to the singlets above,
gives $\nu=0$ for the doublet: in general the Dirac operator will have no
zero modes for this doublet though it may have for specific choices
of the $U(2)$ connection, but even then the zero modes will occur in pairs
of opposite chirality.  The spectrum contains one generation of
the electroweak sector of the standard model, but there is an additional
unwanted doublet of the wrong chirality.

\section{Unitary Grassmannians}

The final source of examples that we wish to discuss is the unitary Grassmannians 
\begin{equation}
{U(n)\over U(k)\times U(n-k)}\cong {SU(n)\over S(U(n-k)\times U(k))}
\end{equation}
of which the complex projective spaces, $k=1$, are special cases.
The first Chern class of the tangent bundle for these space evaluates to $n$, 
\cite{HirzebruchBorel},
and the second Stiefel-Whitney class is $n$ mod $2$---so these spaces admit a spin
structure if and only if $n$ is even.  We shall focus on the particular case
of $n=5$ and $k=2$, this is an interesting 
case because the holonomy group of $SU(5)/S(U(3)\times U(2))$
is precisely that of the standard model, \cite{ORaifeartaigh}.  
This condition dictates that the
Fermions actually sit in representations of $SU(3)\times SU(2)\times U(1)$
in which the generators are traceless---whence $S(U(3)\times U(2))$.
As a matrix manifold $SU(5)/S(U(3)\times U(2))$ was studied in \cite{G2N},
where a star product was explicitly constructed in terms of derivatives.

The Grassmannian $SU(5)/S(U(3)\times U(2))$ has Euler characteristic $\chi=10$ 
and  signature
$\tau=2$.  It is not a spin manifold but a $Spin^c$ structure 
exists.  Taking the bundle $\wedge^{0,*}TX\otimes L^{-q}$, with 
$X$ the Grassmannian and $L$ the generating line bundle,
the Dirac index is calculated in appendix D as
(\ref{eq:Grassmannianq}),
\begin{equation}\nu_{\{q,{\bf 1},{\bf 1}\}}=
{1\over 144}(q+1)(q+2)^2(q+3)^2(q+4),\end{equation}
where the notation $\{q,{\bf 1},{\bf 1}\}$ indicates the $ U(1)\times SU(3)\times SU(2)$
structure of the vector bundle.

The Chern character $ch(L^{-q})$ evaluates to $q$ on a 2-sphere
so the $U(1)$ charge here is $q$ which we shall rescale by a factor of $2$
in order to reproduce the standard model quantum numbers later,
\begin{equation}Y_{\{q,{\bf 1},{\bf 1}\}}=2q.\end{equation}
In particular with $q=0$ gives
$\nu_{\{0,{\bf 1},{\bf 1}\}}=1$, and
interpreting positive chirality as right-handed
gives a single generation of right-handed neutrino, while $q=-1$ gives
$Y_{\{-1,{\bf 1},{\bf 1}\}}=-2$ which would be the right-handed electron but
$\nu_{\{-1,{\bf 1},{\bf 1}\}}=0$.

We can include fundamental $SU(2)$ gauge fields by taking the spinors to transform as a 
doublet and constructing the rank 2 vector bundle $F$ with structure group $U(2)$
associated with the principal bundle
\begin{equation}
\matrix{U(2)&\longrightarrow & U(5)/U(3)\cr & & \downarrow\cr 
&& U(5)/ (U(3)\times U(2))\ .\cr}
\end{equation}
The Dirac index of the bundle $\wedge^{0,*}TX\otimes F\otimes L^{-q}$ is calculated in the appendix and
shown to be (\ref{eq:GrassmannianFq})
\begin{equation} \nu_{\{q,{\bf 1},{\bf 2}\}} = 
{1\over 72}(q+1)(q+2)(q+3)^2(q+4)(q+5).\end{equation}
The first Chern class $c_1(F)$ evaluates to $1$ and the
Chern character $ch( F\otimes L^{-q})$ to 
$2q+1$ on a non-trivial $S^2$
so, dividing by the rank of the bundle, 
the $U(1)$ charge here is $q+{1\over 2}$ which is rescaled by a factor 
of 2 as before to
$Y_{\{q,{\bf 1},{\bf 2}\}}=2q+1$.
In particular $q=0$ gives $Y_{\{0,{\bf 1},{\bf 2}\}}=1$ and
$\nu_{\{0,{\bf 1},{\bf 2}\}}=5$
which we interpret
as copies of the positron--anti-neutrino doublet
$\left(\matrix{
(\overline{\hbox{\Nu V \,}})_{\bf R}\cr (\overline{e})_{\bf R} \cr}\right)$.

Fundamental $SU(3)$ gauge fields can be incorporated by a very similar procedure:
take the spinors to transform as an $SU(3)$
triplet and construct the rank 3 vector bundle $E$ with structure group $U(3)$
associated with the bundle
\begin{equation}\matrix{U(3)&\longrightarrow & U(5)/U(2)\cr & & \downarrow\cr 
&& U(5)/ [U(3)\times U(2)]\ .\cr}
\end{equation}
The index of the bundle $\wedge^{0,*}TX\otimes E\otimes L^{-q}$ 
is (\ref{eq:GrassmannianEq})
\begin{equation} \nu_{\{q,{\bf 3},{\bf 1}\}} = 
{1\over 48}q(q+1)(q+2)(q+3)^2(q+4).\end{equation}
The bundles $E$ and $F$ are related by $E\oplus F=I^5$,
so $c_1(E)$ of $E$ evaluates to $-1$ so integrating  
$ch(E\otimes L^{-q})$ over an $S^2$ gives $3q-1$ 
giving $U(1)$ charge $q-{1\over 3}$ which we rescale by $2$ to give 
$Y_{\{q,{\bf 3},{\bf 1}\}}=2q-{2\over 3}$.

In particular $q=0$ gives the right-handed $d_{\bf R}$
but again
\begin{equation}
\nu_{\{0,{\bf 3},{\bf 1}\}}=0.
\end{equation}
The right-handed $u_{\bf R}$ quarks arises from $q=1$ giving 
$Y_{\{1,{\bf 3},{\bf 1}\}}=4/3$ with index 
\begin{equation}
\nu_{\{1,{\bf 3},{\bf 1}\}}=10.
\end{equation}

The left-handed quarks of the standard model are both $SU(3)$ triplets and $SU(2)$
doublets so the bundle $\wedge^{0,*}TX\otimes E\otimes F \otimes L^{-q}$ is also considered
in the appendix (\ref{eq:GrassmannianEFq}), leading to
\begin{equation}
\nu_{\{q,{\bf 3},{\bf 2}\}} = 
{1\over 24}q(q+2)^2(q+3)(q+4)(q+5).
\end{equation}
The Chern character $ch(E\otimes F\otimes L^{-q})$ integrates to
$6q+1$ on a non-trivial $S^2$, so the $U(1)$ charge is
$q+{1\over 6}$ which is rescaled to 
$Y_{\{q,{\bf 3},{\bf 2}\}}=2q+{1\over 3}$.
The choice $q=0$ leads to the quark doublet, 
$\left(\matrix{u_{\bf L}\cr d_{\bf L}\cr}\right)$, 
with $Y_{\{0,{\bf 3},{\bf 2}\}}=1/3$,
and 
\begin{equation}
\nu_{\{0,{\bf 3},{\bf 2}\}}=0.
\end{equation}

To summarise we can find the standard model charge assignments with 
the unitary Grassmannian $U(5)/\left(U(3)\times U(2)\right)$,
but the indices, and so the multiplicities, are wrong:
\begin{equation} \hbox{\Nu V \,}_{\bf R} \quad \nu_{\{0,{\bf 1},{\bf 1}\}}=1,\qquad
e_{\bf R} \quad\nu_{\{-1,{\bf 1},{\bf 1}\}}=0,\qquad
d_{\bf R} \quad \nu_{\{0,{\bf 3},{\bf 1}\}}=0,\qquad
u_{\bf R}\quad \nu_{\{1,{\bf 3},{\bf 1}\}}=10,\end{equation}
\begin{equation}
\left(\matrix{ (\overline{\hbox{\Nu V \,}})_{\bf R}\cr(\overline{e})_{\bf R}\cr}\right)\quad
\nu_{\{0,{\bf 1},{\bf 2}\}}=5,\qquad
\left(\matrix{u_{\bf L}\cr d_{\bf L}\cr}\right)\quad
\nu_{\{0,{\bf 3},{\bf 2}\}}=0.
\end{equation}
Obviously this is unsatisfactory as it stands: the multiplets with zero index will not
be zero-modes in general and even if they are they will be accompanied with zero-modes
of the opposite handedness but the same hypercharge; also $5$ weak doublets and $10$
right-handed $u$-quark `families' is clearly not compatible with the current
experimental picture.

\section{Conclusions}

We have investigated zero modes of the Dirac operator on various manifolds
which admit finite matrix approximations.  Such spaces are candidates for
finite internal spaces in non-conventional Kaluza-Klein theory, where the internal
space consists of a finite number of points.  In this paper we have
focused on manifolds that do not admit a spin structure, as the inevitable
twisting of bundles that enables spinors to be defined ($Spin^c$ structures)
unavoidably leads to chiral Fermions.  The electroweak sector of the standard
model emerges naturally in this construction from $\CP^2\cong SU(3)/U(2)$:
the gauge group is $U(2)$ and
the usual $Spin^c$ structure gives rise to a neutral singlet which
is identified with the right-handed neutrino while tensoring the standard
$Spin^c$ bundle with the inverse of the canonical line bundle gives another $SU(2)$
singlet with the quantum numbers of the right-handed electron.
The electron-neutrino doublet arises by coupling spinors to a natural
rank 2 bundle which is dual to the generating line bundle---the curvature
associated with this bundle represents a $U(2)$ instanton on $\CP^2$.

The resulting spectrum represents one generation of the electroweak sector
of the standard model.  The smallest non-trivial matrix approximation to
$\CP^2$ is the algebra of $3\times 3$ matrices, \cite{FuzzyCPN}, acting
on a three dimensional complex vector space which carries the fundamental
representation of the isometry group $SU(3)$.  It may be that this
could be interpreted as a horizontal symmetry giving rise to three generations.
Note that the philosophy here is rather different to the usual Kaluza-Klein approach
where the isometry group is identified with the gauge group---here the isometry
group is being identified with a horizontal symmetry group and the holonomy
group is the gauge group.

We have also investigated two other manifolds: a six dimensional manifold with
holonomy group $U(2)$, $Sp(2)/U(2)$; and a twelve dimensional manifold
with holonomy group $SU(3)\times SU(2)\times U(1)$, the unitary Grassmannian
$U(5)/\bigl( U(3)\times U(2) \bigr)$.  These manifolds both admit finite
matrix approximations and neither admits a spin structure.  The former
gives a spectrum containing
the correct charges for the electroweak sector of the standard model,
but the electron-neutrino doublet has zero index: generically
the Dirac operator would have no zero modes corresponding to this doublet.
There may exist particular
connections for which the doublet is a zero mode but this would necessarily be accompanied
by a doublet of the opposite chirality, unless some other mechanism could be invoked
to eliminate it.
The unitary Grassmannian
gives the correct representations and charges for the whole Fermionic
sector of the standard model, but again some multiplets have zero
index and here the multiplicities are different for the multiplets with non-zero index.

A number of questions remain to be addressed.  Obviously it is of interest
to look further for other manifolds that might give a better fit to the
standard model spectrum with this approach.  One possibility is to consider
manifolds that admit spinors directly, without the necessity of a $Spin^c$
structure. After all it is only the electroweak sector of the standard
model that requires different representations for right and left-handed particles
and, as we have seen, this can be obtained from $\CP^2$.  QCD does not require
any such asymmetry and so could arise more directly, without the introduction
of chirally asymmetric Fermions.  Indeed we hope to show elsewhere
that this is indeed the case \cite{SMCPN}.

There is also the question of the Higgs sector of the standard model, 
which we have not addressed here.
In Connes' approach to the standard model, the Higgs field is associated
with two `copies' of space time, which could be viewed as coming from an internal
space consisting of two points acted on by the $SU(2)$ symmetry of weak interactions.
This looks very much like a two dimensional vector space acted on by a matrix
approximation to the 2-sphere.  
Grand unified models could also be investigated with the techniques
used here.

Lastly we have assumed that the usual differential-geometric analysis of
the Atiyah-Singer index theorem on continuous manifolds will carry over
to finite matrix approximations without change.  While this seems reasonable to us
there is certainly no proof that it is true in general, but this
would require a much more involved investigation than is presented here.

\appendix

\section{Spin and spin$^c$ structures on a manifold}
\label{ap:spinc}
In this appendix we provide details for the calculations presented in 
the main text; useful references for this material are 
Borel and Hirzebruch  \cite{HirzebruchBorel}, Michelson and Lawson \cite{LawMich} and  
Bott and Tu \cite{Bott+Tu}.
\par 
This section describes, in  brief, what is involved for a manifold $X$ to admit 
 spinors---$X$ is then said to have a {\it spin structure} or to 
be a {\it spin manifold}---and  
failing that, we describe how $X$ can have what is called a {\it spin$^c$ structure}. 
A manifold $X$ has a {\it spin$^c$ structure} when it 
 admits  a certain  pair consisting of a 
$U(1)$ connection and a `local spinor'---$X$ is then said  be a {\it spin$^c$ manifold}. 
Spin manifolds are automatically spin$^c$ manifolds but the converse is false. 
\par
If an $n$ dimensional   
manifold $X$ (compact and closed in this discussion) is a spin 
manifold then its tangent bundle 
\beq
TX 
\enq
whose principal bundle we denote by 
\beq
P_{SO(n)}(X)
\enq
has structure group $SO(n)$. 
Sections of $TX$ are then vectors on $X$.
\par
The fact  that $X$ is  spin means that $TX$ 
possesses a lifting of its structure 
group from the group $SO(n)$ to the group $Spin(n)$. 
Such a 
lifting, which need not be unique, constitutes a choice of spin 
structure on $X$. This lifting induces from $P_{SO(n)}(X)$ a 
$Spin(n)$ principal bundle
\beq
P_{Spin(n)}(X)
\enq
on $X$; also induced from $TX$, and associated to $P_{Spin(n)}(X)$, is 
the bundle of spinors 
\beq
S(X) 
\enq
over $X$. Finally  sections of $S(X)$ are  called spinors. 
\par
The existence of  this lifting requires topological obstructions to 
vanish namely that the first  two Stiefel--Whitney classes of $TX$
should vanish i.e.
\beq
w_1(X)=0,\qquad w_2(X)=0.
\enq
The vanishing of $w_1(X)$ just guarantees that $X$ is orientable 
and allows us to distinguish clockwise and anti-clockwise rotations; but 
the vanishing of $w_2(X)$ is needed to make the double covering of 
of $SO(n)$ by $Spin(n)$ work globally. 
\par
If $X$ is orientable but 
\beq
w_2(X)\not=0 
\enq
then global spinors do not exist and $X$ is not a spin manifold.
\par 
When $X$ is not a spin manifold the situation can be saved if $X$ admits 
a generalisation of a spin structure known as a spin$^c$ structure; moreover this is 
quite a natural structure if $X$ is a complex manifold, though $X$ 
does not need to be complex. An orientable $X$ admits a  spin$^c$ structure if 
$w_2(X)$ is the reduction ${\rm mod\,}2$ of 
an integral cohomology class in $H^2(X;{\bf Z})$. This is guaranteed for complex 
manifolds 
since their Chern classes determine their Stiefel--Whitney classes via the relation  
\beq
w_2(X)=c_1(X){\rm \,mod\,}2.
\enq
\par
The underlying mechanism which makes a spin$^c$ structure work is easy to expose when 
$X$ is complex. Suppose then that $X$ has a K\"ahler metric and is complex with complex 
dimension $n$;  let the canonical line bundle of $X$ be $K$ so that 
\beq
K=\wedge^n T^*X.
\enq
Recall also for later use that $c_1(K)=-c_1(X)$.  
Now suppose first that $X$ is a spin manifold so that 
\beq
w_2(X)=0\Rightarrow c_1(X) \hbox{ is even} 
\enq
and that the spinor bundle $S(X)$ does exist; this in turn means that the canonical bundle 
$K$ has square roots: a choice of square root is a spin structure. Now   
consider the bundle $\wedge^{0,*}TX$ of all forms of type $(0,s)$---i.e. anti-holomorphic 
$s$ forms---so we have
\beq
\wedge^{0,*}TX={\textstyle\bigoplus\limits_s}\wedge^{0,s}\longbar{T^*X}.
\enq
The the spinor bundle $S(X)$ is obtained by just tensoring $\wedge^{0,*}TX$ with 
$K^{1/2}$ i.e.
\beq 
S(X)=\wedge^{0,*}TX\otimes K^{1/2}.
\enq
Hence spinors are $K^{1/2}$-valued $(0,s)$ forms---which we denote by 
$\Omega^s(K^{1/2})$---and are sections of $S(X)$.
The full self-adjoint Dirac $\Dirac$ operator is loosely 
$\bar\partial_{K^{1/2}}+\bar\partial_{K^{1/2}}^*$, the $\bar \partial$ 
operator acting on sections of 
$S(X)$; the chiral Dirac operator is denoted by $\dirac$ and this setup now gives us 
 \begin{eqnarray}
\Dirac&=&\left(\matrix{0&\dirac\cr \dirac^*&0\cr}\right)\qquad\Dirac\equiv\sqrt{2}\left(\bar\partial_{K^{1/2}}+\bar\partial_{K^{1/2}}^*\right)\\
\dirac&:& {\textstyle\bigoplus\limits_p}\,\Omega ^{2p}(K^{1/2})\longrightarrow {\textstyle\bigoplus\limits_p}\,\Omega^{2p+1}(K^{1/2})\\
\dirac^*&:&{\textstyle\bigoplus\limits_p}\,\Omega^{2p+1}(K^{1/2})\longrightarrow
{\textstyle\bigoplus\limits_p}\,\Omega ^{2p}(K^{1/2}).
\end{eqnarray} 
The chirality of a spinor now corresponds to its parity as a form. 
\par
All the above was for the case when $X$ is spin. Now suppose that $X$ is {\it not} spin
then we see that
\begin{eqnarray}
w_2(X)\not=0\Rightarrow &c_1(X)&=-c_1(K)\hbox{ is odd}\\
            \Rightarrow &K^{1/2}&\hbox{does not exist}\\                   
            \Rightarrow &S(X)&\hbox{does not exist.}
\end{eqnarray}
But, though $S(X)=\wedge^{0,*}TX\otimes K^{1/2}$ does not exist, the bundle 
$\wedge^{0,*}TX$ clearly does: 
 this is the spin$^c$ bundle which we denote by $S^c(X)$ so that
\beq
S^c(X)=\wedge^{0,*}TX.
\enq
Now if we abuse notation temporarily and write down the tensor product of 
the two non-existent bundles $S(X)$ and $K^{-1/2}$ we get the spin$^c$ bundle $S^c(X)$ since we can write  
\beq
S^c(X)=S(X)\otimes K^{-1/2}\hbox{ (`locally')}
\enq
and this bundle $S^c(X)$ does exist globally even though its factors do not. The point 
is that the factors do exist locally and  the failure of one factor to behave properly 
(under, for example parallel transport round a closed loop) is  
compensated for by a failure of the other; this mechanism  renders the 
`product' well defined. This picture of $S^c(X)$ makes it clear at once that the generalised  spinors of a  spin$^c$ structure are also coupled to a local $U(1)$ connection. 
\par
Finally, as we are  interested  in chiral Fermions, we want to point out that the 
Dirac operator exists for generalised spinors and it is natural to denote it by 
\beq
\dirac_{K^{-1/2}}.
\enq
Just as a spin structure need not be unique nor need a spin$^c$ structure: one can tensor the `bundle' 
$K^{-1/2}$ by any other genuine line bundle: all our manifolds $X$ will have one dimensional 
$H^2(X;{\bf Z})$ so that there is a `smallest' line bundle $L$  defined by requiring
\beq
c_1(L)=-1.
\enq
We shall call $L$ the {\it generating line bundle} and, in each case, 
$K$ will be some power of $L$---this power will be odd if $X$ is not 
a spin manifold---so that
\beq
K=L^m,\; m\in{\bf Z}.
\enq 
Hence a general spin$^c$ structure will have the spin$^c$ bundle
\beq
\wedge^{0,*}TX\otimes L^{-q} =S^c(X)\otimes L^{-q},\quad q\in{\bf Z}
\enq
(the minus sign in the exponent is for later convenience).
When $q=0$ we have the {\it canonical} spin$^c$ structure; there is also a dependence of the spin$^c$ structure on an element of 
$H^1(X; {\bf Z}_2)$ but our examples have $H^1(X;{\bf Z})=0$ so 
we do not need to consider this. 
\par
If we use the fact that $K=L^m$ then
the corresponding Dirac operator then becomes $\dirac_{L^{-(q+m/2)}}$ which we shall neaten 
up slightly by writing it as 
\beq
\dirac_{L^p}\quad\hbox{where }p=-q-m/2.
\enq
\par
There is also  an index formula for 
 the zero modes of $\dirac_{L^p}$ which involves the usual $\hat A$ genus of $X$ and the 
`Chern class' of the  line bundle $L^p$.
Let $\dirac_{L^p}$ denote the Dirac operator coupled 
to $L^p$ then its index is given by 
\footnote{We could equally have used instead the formula for 
$\indx(\bar\partial_{L^p})$ which would have involved $ch\,(L)$ and the Todd class $td(X)$. 
In fact this realisation of the Dirac operator  as 
$\bar\partial_{L^p}$ enables one to easily understand why  
$\indx(\dirac_{L^{-q-m/2}})$ is equal to unity for $q=0$:  it is because, when 
$q=0$, $\indx(\bar\partial_{L^{-m/2}})$ gives the arithmetic genus 
$\sum(-1)^s h^{0,s}$ 
of the complex manifold $X$ where the Hodge number $h^{r,s}$ denotes 
the dimension  of the space of holomorphic forms of type $(r,s)$. Now for the manifolds $X$  
we consider in this paper the only holomorphic forms are of type $(s,s)$ 
a fact which reduces the arithmetic genus to 
$h^{0,0}$ which is trivially unity.}
\begin{eqnarray}
\indx(\dirac_{L^p}) & = & \hbox{\rm ch\,}(L^p)\hat A(X)[X] \\
                       & = &\exp\left[p c_1(L)\right]\hat A(X)[X].
\end{eqnarray}
We will also need the case where the Dirac operator is further coupled to 
a second vector bundle $E$ of rank possibly greater than one; in this case the requisite index formula is

\begin{eqnarray}
\indx(\dirac_{L^p\otimes E}) & = & \hbox{\rm ch\,}(L^p \otimes E)\hat A(X)[X] \\
                       & = &\hbox{\rm ch\,}(E)\exp\left[p c_1(L)\right]\hat A(X)[X],\;(p=-q-m/2). 
\end{eqnarray}
\par
In the next  section we treat an actual  spin$^c$ example in 
four dimensions. 

\section{Cohomology and spin$^c$ on $\CP^2$}
\label{ap:CP2}
On $\CP^2$ the Chern class is \cite{Bott+Tu}
\begin{equation}
c(\CP^2)=1-3c_1(L)+3c_1^2(L)
\end{equation} where the generating line bundle $L$
has $c(L)=1+c_1(L)$ with $-c_1(L)$ generated by the K\"ahler 2-form. 
The Euler characteristic is $3$ so $c_1^2(L)[\CP^2]=1$
and in this case $c_1(\CP^2)=-3c_1(L)$ so $m=3$.
Since the coefficient of $c_1(L)$ is odd $w_2\ne 0$ and $\CP^2$ does not admit
a spin structure.
The index of the Dirac operator coupled to $L^p$ is
\begin{eqnarray}
\indx(\dirac_{L^p}) & = & \hbox{\rm ch\,}(L^p)\hat A(X)[X] 
                       = \exp\left[p c_1(L)\right]\hat A(X)[X]\\
			& = & 
         \left(1+pc_1(L)+{1\over 2}p^2c_1^2(L)\right)\left(1-{p_1(X)\over 24}\right)[X],\quad X=\CP^2\\
& = & {1\over 8}(4p^2-1),\;
\end{eqnarray}
where $p_1$ is the Pontrjagin class and  $p_1(X)=c_1^2(X)-2c_2(X)=3c_1^2(L)$ on $\CP^2$.
This index is integral for half-integral $p$ and, setting $p=-q-3/2$, 
we obtain
\begin{eqnarray}
\indx(\dirac_{L^p}) = {1\over 2}(q+2)(q+1).
\label{eq:CPq}
\end{eqnarray}

We can define a non-trivial rank 2 bundle $F$ over $\CP^2$ with structure group
$U(2)$ by $F\oplus L\cong I^3$ where $I^3$ is the trivial rank 3 bundle.
Then $c(F)\,c(L)=1$ so $c_1(F)=-c_1(L)$ and $c_2(F)=c_1^2(L)$;
tensoring this with $p$ copies of the generating line bundle $L$ 
then gives, for the  Chern character,  
\begin{eqnarray}
  ch(L^p\otimes F)&=&ch(L^p)ch(F)\\
    &=&\left(1+pc_1(L)+{1\over 2}p^2c_1^2(L)+\cdots\right)
\left(2-c_1(L)-{1\over 2}c_1^2(L)+\cdots\right)\\
    &=& 2+(2p-1)c_1(L)+\left(p^2-p-{1\over 2}\right)c_1^2(L)+\cdots,
\end{eqnarray}
leading to 
\begin{eqnarray}
\indx(\dirac_{L^p\otimes F}) & = & \hbox{\rm ch\,}(F\otimes L^p)
\hat A(X)[X] 
                       = \exp\left[p c_1(L)\right]\hbox{\rm ch\,}(F)\hat A(X)[X]\\
			& = & 
\left(2+(2p-1)c_1(L)+\left(p^2-p-{1\over 2}\right)c_1^2(L)\right)\left(1-{p_1\over 24}\right)[X]
\nonumber\\
& = & {1\over 4}(2p-3)(2p+1), \quad X=\CP^2\\
&=&(q+1)(q+3), \qquad\hbox{again using} \ p=-q-3/2.
\label{eq:CPFq}
\end{eqnarray}

\section{Cohomology and spin$^c$ in six dimensions}
\label{ap:Sp2}
In this section $X$ is the complex manifold given by
\beq
X={Sp(2)\over U(2)}
\enq
whose real dimension is $6$. The cohomology ring of $X$ is generated by the even 
dimensional classes  $\sigma_1\in H^2(X;{\bf Z})$ and $\sigma_2\in H^4(X;{\bf Z})$ 
subject to the single relation
\beq
\sigma_1^2=2\sigma_2.
\enq
Now $X$ is not a spin$^c$ manifold because we can compute that
\begin{eqnarray}
c(X)&=&1+c_1(X)+c_2(X)+c_3(X)\\
    &=&1+3\sigma_1+8\sigma_2+4\sigma_1\sigma_2\\
    \Rightarrow c_1(X)&=&3\sigma_1=-3 c_1(L).
\end{eqnarray}
We note that $\sigma_1$ generates $H^2(X;{\bf Z})$ and so deduce that $c_1(X)$ is odd and so
\beq
w_2(X)\not=0 \Rightarrow X \hbox{ is not spin}.
\enq
We also see that
\beq
K=L^3
\enq
so that the integer $m$ of appendix \ref{ap:spinc}  is equal to $3$.
\par
The index of the Dirac operator $\dirac_{L^p}$ can now be computed from the expansions of 
$ch\,(L^p)$ and $\hat A(X)$ giving us the formula
\begin{eqnarray}
\indx(\dirac_{L^p})&=&\left(1+p c_1(L)+{1\over2}p^2 c_1^2(L)+
\cdots\right)\left(1-{p_1(X)\over 24}+\cdots\right)[X]\\
&=&\left(- pc_1(L){p_1(X)\over 24}+{1\over 3!}p^3 c_1^3(L)\right)[X].
\end{eqnarray}
But we can calculate that
\begin{eqnarray}
p_1(X)&=&c_1^2(X)-2c_2(X)\\
      &=&9\sigma_1^2-16\sigma_2,
\end{eqnarray}
with $\sigma_1=-c_1(L)$.
Hence we find that
\begin{eqnarray}
\indx(\dirac_{L^p})&=&\left({p\over 24}\sigma_1(9\sigma_1^2-16\sigma_2)-{1\over 3!}p^3\sigma_1^3\right)[X]\\
&=&-(4p^3-p){\sigma_1^3\over 24}[X]\\
&=&-{1\over 12}(4 p^3-p)=-{1\over12}p(2p-1)(2p+1),
\end{eqnarray}
where we have used the Gauss--Bonnet theorem which says that
\begin{eqnarray}
c_3(X)[X]&=&\chi(X)\\
         &=&4=2\sigma_1^3[X]
\end{eqnarray}
to deduce that $\sigma_1^3[X]=2$.  
Before finishing we should check that the index is integral. Recall that
\beq
p=-q-m/2,\quad q\in{\bf Z},\;m=3
\enq
This fact immediately gives us the formula
\beq
\indx(\dirac_{L^p})={1\over 6}(2q+3)(q+1)(q+2),\;q\in{\bf Z}
\label{eq:Spq}
\enq
and this is easily checked to give an integer index for integral $q$ as it should.
\par
If we tensor product with a further rank 2 bundle $F$, with $c_1(F)=\sigma_1$ 
then we find that
\begin{eqnarray}
\indx(\dirac_{L^p\otimes F})&=&ch(L^p\otimes F)\hat A(X)[X]\\
&=&ch(L^p)\,ch(F)\hat A(X)[X]\\
                            &=&-{1\over 12}(2p+3)(2p-1)(2p+1)\\
                &=&{2\over3}q(q+1)(q+2),\quad p=-q-3/2,\;q\in{\bf Z}
\label{eq:SpqF}
\end{eqnarray}
and again this gives an integral index.

\section{Cohomology and generalised spinors for a $12$
 dimensional Grassmannian.}
\label{ap:G2N}
In this section $X$ is the $12$ dimensional Grassmannian given by 
\beq
X={U(5)\over U(3)\times U(2)}.
\enq
$X$ is a perfectly standard complex manifold (of complex dimension $6$) and its cohomology ring
$H^*(X;{\bf Z})$ has $3$ generators
\beq
\sigma_i\in H^{2i}(X;{\bf Z}),\;i=1,2,3
\enq
which obey the single relation
\beq
\sigma_3=2\sigma_1\sigma_2-\sigma_1^3.
\label{eq:srelation}
\enq

Its Chern class is given by
\begin{eqnarray}
c(X)&=&(1+c_1(X)+c_2(X)+c_3(X)+c_4(X)+c_5(X)+c_6(X))\\
&=&(1-5\sigma_1+12\sigma_1^2-15\sigma_1^3+8\sigma_1^4+2\sigma_1^2\sigma_2+7\sigma_2^2+
4\sigma_1^5-25\sigma_1\sigma_2^2\\
&&-29\sigma_1^6+7\sigma_1^2\sigma_2^2+56\sigma_1^4\sigma_2-27\sigma_2^3)
\end{eqnarray}
from which we see that
\beq
c_1(X)=-5\sigma_1
\enq
and hence we deduce, as we did in the previous section,  that 
\beq
w_2(X)\not=0
\enq
and so $X$ is not spin. 
\par
We now pass to the spin$^c$ bundle $S^c(X)$ and to the calculation of the index of its Dirac operator
$\dirac_{L^p}$ where $L$ is the generating line bundle as it was in the previous section. 
But this time we need the fact that $\sigma_1$ is actually a {\it negative generator} of 
$H^2(X,{\bf Z})$ with our orientation conventions  and so we have
\begin{eqnarray}
c_1(X)&=&-5\sigma_1,\quad \sigma_1 \ \hbox {negative,}\quad \sigma_1=c_1(L)\\
\Rightarrow K&=&L^5\quad (m=5)\\
\hat A(X)&=&\left(1-{p_1(X)\over 24}+{1\over 5760}(7p_1^2(X)-4p_2(X))\right.\\
&&\left.-{1\over 2^{10} \cdot 945}(16p_3(X)-44p_1(X)p_2(X)+31p_1^3(X))+\cdots\right)
\end{eqnarray}
as well as
\begin{eqnarray}
p_1(X)&=&c_1^2(X)-2c_2(X)=\sigma_1^2+2\sigma_2\\
p_2(X)&=&-2c_1(X)c_3(X)+c_2^2(X)+2c_4(X)=10\sigma_1^4  - 20 \sigma_1^2\sigma_2 + 15\sigma_2^2\\
p_3(X)&=&2c_1(X)c_5(X)-2c_2(X)c_4(X)+c_3^2(X)-2c_6(X)\\
      &=&51\sigma_1^6+72\sigma_1^2\sigma_2^2-144\sigma_1^4\sigma_2+68\sigma_2^3.
\end{eqnarray}
This information allows to compute that
\begin{eqnarray}
\indx(\dirac_{L^p}) &=& \exp\left[p c_1(L)\right]\hat A(X)[X]\\
                    &=&-{1\over 60480}\sigma_2^3[X]-\left({41\over 15120}+{1\over 360}p^2\right)\sigma_1^2\sigma_2^2[X]\\ 
&&+\left({353\over161280}+{3\over 320}p^2-{1\over 288}p^4\right)\sigma_1^4\sigma_2[X]\\
&&+\left(-{407\over 967680}-{11\over 3840}p^2-{1\over 576} p^4+{1\over 720}p^6\right)\sigma_1^6[X].
\end{eqnarray}

Now use the cohomology generators and the fact that $X$ 
clearly has Euler characteristic $10$ we discover that
\begin{eqnarray}
\sigma_2^3[X]=1\\
\sigma_1^2\sigma_2^2[X]=2\\
 \sigma_1^4\sigma_2[X]=3\\ 
\sigma_1^6[X]=5.
\end{eqnarray}
This all gives the formulae
\begin{eqnarray}
\indx(\dirac_{L^p}) &=&-{1\over 1024}+{19\over 2304}p^2-{11\over 576}p^4 +{1\over 144}p^6\\
&=&{1\over 9.2^{10}}(4p^2-9)(4p^2-1)^2\\
                    &=&{1\over 144}(q+1)(q+2)^2(q+3)^2(q+4),\quad\hbox{ using } p=-q-5/2, 
\label{eq:Grassmannianq} 
\end{eqnarray}
and this index is an integer for integral $q$ as required.
\par
We shall finish  by calculating  the index when we couple the Dirac operator 
to some higher rank bundles. 
We shall give the results for two bundles $E$ and $F$ which are naturally associated to $X$ and also for the tensor product $E\otimes F$. 
\par
Let $E$ be the rank $3$ vector bundle over 
\beq
X={U(5)\over U(3)\times U(2)}
\enq
whose fibre over a point $x\in X$ is the $3$-plane $x$ itself. This describes the bundle $E$.
Now consider the product rank $5$ bundle $X\times {\bf C^5}$ then $F$ is the rank $2$ bundle created
by forming the quotient
\beq
{X\times {\bf C^5}\over E}.
\enq
The bundles $E$ and $F$ satisfy 
\beq
E\oplus F \cong I^5  
\enq
where $I^5$ is a trivial rank $5$ bundle  and it is not difficult to work out that
\begin{eqnarray}
&&c(E)c(F)=1\\
\hbox{i.e. }&&(1+c_1(E)+c_2(E)+c_3(E))(1+c_1(F)+c_2(F))=1\\
&&ch(E)+ch(F)=5.\label{eq:EFconstraint}
\end{eqnarray}  
In fact equation (\ref{eq:EFconstraint}) can be used to derive 
the relation (\ref{eq:srelation})
since the classes $\sigma_i$ are just the classes $c_i(E)$ and so this 
allows all of
$c(E)$ and $c(F)$ to be expressed in terms of the $\sigma_i$.

 The Chern characters of $E$ and $F$ are given by
\begin{eqnarray*}
ch(E)&=&
3+c_1(E)+{1\over2}\left(c_1^2(E)-2c_2(E)\right)+{1\over 3!}\left(c_1^3(E)-3c_1(E)c_2(E)+3c_3(E)\right)+
{1\over 4!}\left(c_1^4(E)\right.\\
&&\left.-4c_1^2(E)c_2(E)+4c_1(E)c_3(E)+2c_2^2(E)\right)+
{1\over 5!}\left(c_1^5(E)
-5c_1^3(E)c_2(E)+5c_1^2(E)c_3(E)\right.\\
&&\left.+5c_1(E)c_2^2(E)-5c_2(E)c_3(E)\right)+
{1\over 6!}\left(c_1^6(E)-6c_1^4(E)c_2(E)+6c_1^3(E)c_3(E)\right.\\
&&\left.+9c_1^2(E)c_2^2(E)
-12c_1(E)c_2(E)c_3(E)-2c_2^3(E)+3c_3^2(E)\right)\\
ch(F)&=&5-ch(E).
\end{eqnarray*}  
 
\par
Now we can calculate the index of the appropriate Dirac operators: Forming the product 
$L^p\otimes E$ we have 
\begin{eqnarray}
\indx(\dirac_{L^p\otimes E}) & = & \hbox{\rm ch\,}(L^p \otimes E)\hat A(X)[X] \\
                     & = &\hbox{\rm ch\,}(E)\exp\left[p c_1(L)\right]\hat A(X)[X],
\end{eqnarray}
 and we find that 
\begin{eqnarray}
\indx(\dirac_{L^p\otimes E})&=&-{15\over 1024}+{3\over 128}p+{59\over 768}p^2-{5\over 48}p^3
-{5\over 64}p^4+{1\over 24}p^5+{1\over 48}p^6\nonumber\\
&=& {1\over 3.2^{10}}(2p+5)(2p-1)(4p^2-9)(4p^2-1)\\
                            &=&{1\over 48}q(q+1)(q+2)(q+3)^2(q+4),\;(p=-q-5/2)
\label{eq:GrassmannianEq}
\end{eqnarray}
and for the product $L^p\otimes F$ 
\begin{eqnarray}
\indx(\dirac_{L^p\otimes F})&=&{5\over 512}-{3\over 128}p-{41\over 1152}p^2+{5\over 48}p^3-{5\over 288}p^4-{1\over 24}p^5+{1\over 72}p^6 \nonumber \\
&=& {1\over 9.2^9}(2p-5)(2p-1)(4p^2-9)(4p^2-1)
\\
&=&{1\over 72}(q+1)(q+2)(q+3)^2(q+4)(q+5),\;(p=-q-{5\over 2}).
\label{eq:GrassmannianFq}
\end{eqnarray}
Finally for the bundle $L^p\otimes E\otimes F$ we have 
\begin{eqnarray}
\indx(\dirac_{L^p\otimes E\otimes F})&=&-{25\over512}-
{25\over 384}p+{103\over 384}p^2+{13\over 48}p^3-{29\over 96}p^4-{1\over 24}p^5+{1\over 24}p^6
\nonumber \\
&=&{1\over 3.2^9}(4p^2-25)(2p-3)(4p^2-1)(2p+1)
\\
&=&{1\over 24}q(q+2)^2(q+3)(q+4)(q+5),\;(p=-q-5/2)
\label{eq:GrassmannianEFq}
\end{eqnarray}
and in each case one can verify that the index is an integer.


\begin{thebibliography}{99}

\bibitem{Szabo} R.R.~Szabo, {\sl Quantum Field Theory on Noncommutative Spaces}, {\tt hep-th/0109162};
J.M.~Garcia-Bond\'ia, {\sl Noncommutative Geometry and Fundamental Interactions}, 
{\tt hep-th/0206006};
J.C.~V\'arilly, {\sl The Interface of Noncommutative Geometry and Physics}, {\tt hep-th/0206007}

\bibitem{ConnesLott} A.~Connes and J.~Lott
{\it Nucl. Phys. B (Proc. Suppl.)} {\bf 18}, (1990), 29

\bibitem{Connes} {\sl Non-commutative Geometry} A.~Connes, (1994), Academic Press

\bibitem{Perelomov} Perelomov, {\sl Generalised Coherent States} (1986) Springer

\bibitem{Berezin} F.~A.~Berezin {\it Commun. Math. Phys.} {\bf 40}, (1975), 153;
{\it Math. USSR, Izv.} {\bf 9}, (1975), 341.

\bibitem{MadoreS2} J.~Madore, {\it Class, Quan. Grav.} {\bf 9}, (1992), 69 

\bibitem{PeterS2} P.~Pre\v{s}najder, {\it J.Math.Phys.} 41 (2000) 2789, {\tt hep-th/9912050}

\bibitem{BalnVaidya} A.P.~Balachandran and S.~Vaidya, {\it Int. J. Mod. Phys.} {\bf A16}, (2001), 17 {\tt hep-th/9910129}

\bibitem{FuzzyCPN} A.P.~Balachandran, B.P.~Dolan, J.~Lee, X.~Martin and D.~O'Connor
{\sl Fuzzy complex projective spaces and their star-products}  {\tt hep-th/0107099}.

\bibitem{G2N} B.P.~Dolan and O.~Jahn, {\sl Fuzzy Complex Grassmannian Spaces and their Star Products}, {\tt hep-th/0111020}

\bibitem{Bordemann} M.~Bordemann, M.~Brischle, C.~Emmrich and S.~Waldmann,
{\it J.~Math.~Phys.} {\bf 37}, (1996), 6311, {\tt q-alg/9512019};
{\it Lett.~Math.~Phys.} {\bf 36}, (1996), 357, {\tt q-alg/9503004}

\bibitem{Schirmer} J.~Schirmer  {\sl A star product for complex Grassmann manifolds}, {\tt q-alg/9709021}

\bibitem{Madore} J.~Madore, {\it Phys. Rev.} {\bf D 41}, (1990), 3709

\bibitem{Grosse} H.~Grosse and  A.~Strohmaier, {\it Lett. Math. Phys.} {\bf 48}, 163, (1999)
{\tt hep-th/9902138}

\bibitem{BalCP2} G.~Alexanian, A.P.~Balachandran, G.~Immirzi and B.~Idri, {\it J. Geom. Phys.} 
{\bf 42}, (2002), 28, {\tt hep-th/0103023}

\bibitem{ShelterIslandII} E.~Witten, in {\sl Proceedings of the 1983 Shelter Island Conference
on Quantum Field Theory and the Fundamental Problems of Physics}, Ed. R.~Jackiw, N.N.~Khuri,
S.~Weinberg and E.~Witten, MIT Press, (1985)

\bibitem{SK} S.~Fukuda et al. Super-Kamiokande collaboration 
{\it Phys. Rev. Lett.} {\bf 86}, (2001), 5651, {\tt hep-ex/0103032};
{\it Phys. Rev. Lett.} {\bf 86}, (2001), 5656, {\tt hep-ex/0103033}

\bibitem{SNO}  Q.R.~Ahmed et al. SNO collaboration {\it Phys. Rev. Lett.} {\bf 87} (2001) 071301; 
{\tt nucl-ex/0204008}; {\tt nucl-ex/0204009} 

\bibitem{Peter} P.~Presnajder, private communication.

\bibitem{ORaifeartaigh} L.~O'Raifeartaigh, {\sl Group Structure of Gauge Theories}, (1987), CUP

%\bibitem{} {\sl Forces From Connes' Geometry} T.~Sch\"uker, {\tt hep-th/0111236}

%\bibitem{} P.~Martinetti and R.~Wulkenhaar, {\it J. Math. Phys} {\bf 43} (2002) 182, 
%{\tt hep-th/0104108}

%\bibitem{Porteous} I.R.~Porteous {\sl Topological Geometry}, 2nd Edition, CUP, (1981),
%chapter 13

\bibitem{HawkingPope} S.W.~Hawking and C.~Pope, {\it Phys. Lett.} {\bf 73B}, 42, (1978)

\bibitem{SMCPN} B.P.~Dolan and C.~Nash, 
{\sl The Standard Model Fermion Spectrum From Complex Projective spaces }
{\tt hep-th/0207078}

\bibitem{HirzebruchBorel} A.~Borel and F.~Hirzebruch
{\sl Characteristic classes and homogeneous spaces} \uppercase{i}, 
{\it Amer. Jour. Math.,} {\bf 80}, 458--538, (1958). 

\bibitem{LawMich} H.~Lawson and M.-L.~Michelsohn
{\sl Spin Geometry}, Princeton 
University Press, (1989).

\bibitem{Bott+Tu} R.~Bott and L.W.~Tu {\sl Differential Forms in Algebraic Topology},  
Springer-Verlag, New York, (1982).


\end{thebibliography}
\end{document}